\documentclass[pdflatex,sn-mathphys-num]{sn-jnl}

\usepackage{graphicx}%
\usepackage{amsmath,amssymb,amsfonts}%
\usepackage{amsthm}%
\usepackage{mathrsfs}%
\usepackage[title]{appendix}%
\usepackage{xcolor}%
\usepackage{textcomp}%
\usepackage{manyfoot}%
\usepackage{booktabs}%
\usepackage{algorithm}%
\usepackage{algorithmicx}%
\usepackage{algpseudocode}%
\usepackage{listings}%
\usepackage{stackengine}

\usepackage{array}
\usepackage[caption=false,font=normalsize,labelfont=rm,textfont=rm]{subfig}
\usepackage{stfloats}
\usepackage{verbatim}
\usepackage{tikz}
\usepackage{circledsteps}
\usepackage{makecell}
\usepackage{wrapfig}
\usepackage{threeparttable, rotating, multirow, booktabs, tabularx, graphicx}
\usepackage{xurl}

\usepackage{tikz}

\newcommand{\myCircled}[2][]{%
\ifmmode
  \tikz[baseline=(char.base)]{
    \node[shape=circle,draw,fill=black,text=white,inner sep=1pt,#1] (char) {$\mathsf{#2}$};
  }%
\else
  \tikz[baseline=(char.base)]{
    \node[shape=circle,draw,fill=black,text=white,inner sep=1pt,#1] (char) {\sffamily #2};
  }%
\fi
}

\newcommand{\myCircledWhite}[2][]{%
\ifmmode
  \tikz[baseline=(char.base),scale=#1]{
    \node[shape=circle,draw,fill=white,text=black,inner sep=0.5pt] (char) {$\mathsf{#2}$};
  }%
\else
  \tikz[baseline=(char.base),scale=#1]{
    \node[shape=circle,draw,fill=white,text=black,inner sep=0.5pt] (char) {\sffamily #2};
  }%
\fi
}

\theoremstyle{thmstyleone}%
\theoremstyle{thmstyletwo}%

\theoremstyle{thmstylethree}%

\begin{document}

\title[Article Title]{In-situ Indexing via Memristive Content-Addressable Memory}


\author[1,2]{\fnm{Bing} \sur{Wu}}\email{wubin200@hust.edu.cn}
\author[3]{\fnm{Xueliang} \sur{Wei}}\email{xueliang\_wei@hbut.edu.cn}
\author[1,2]{\fnm{Shiyi} \sur{Song}}\email{songshiyi@hust.edu.cn}
\author[1,2]{\fnm{Yibo} \sur{Liu}}\email{yiboliu@hust.edu.cn}
\author[1,2]{\fnm{Jinpeng} \sur{Liu}}\email{jinpengliu98@hust.edu.cn}


\author*[1,2]{\fnm{Wei} \sur{Tong}}\email{tongwei@hust.edu.cn}
\author[1]{\fnm{Hao} \sur{Tong}}\email{tonghao@hust.edu.cn}
\author[1,2]{\fnm{Yuchong} \sur{Hu}}\email{yuchonghu@hust.edu.cn}
\author*[1,2]{\fnm{Dan} \sur{Feng}}\email{dfeng@hust.edu.cn}

\affil[1]{\orgname{Huazhong University of Science and Technology}, \orgaddress{\city{Wuhan}, \postcode{430074}, \state{Hubei}, \country{China}}}

\affil[2]{\orgdiv{Key Laboratory of Information Storage System \& Engineering Research Center for Data System and Technology}, \orgname{Ministry of Education}, \country{China}}

\affil[3]{\orgname{Hubei University of Technology}, \orgaddress{\city{Wuhan}, \postcode{430068}, \state{Hubei}, \country{China}}}


\abstract{
Processing-in-Memory (PIM) is a proven paradigm for overcoming the ``memory wall". 
However, while data indexing is severely bottlenecked by this same wall, it remains unclear how indexing can effectively benefit from PIM's unique capabilities. 
We present PATH, an in-situ indexing architecture that bridges this gap by leveraging the massive parallelism and inherent data-movement of PIMs. Specifically, we first reformulate the fundamental indexing operations, namely Insert, Search, Update, and Delete, into highly parallel in-situ content-addressable memory operations executed directly within memory arrays. 
Taking hash indexes as a typical case, we elaborate how PATH breaks the inherent trade-off among memory accesses, load factor, and process latency in conventional hashing schemes. 
By adopting ultra-large logical buckets and in-memory moving, PATH virtually eliminates the cost of hash collision resolution and significantly reduces resizing overhead. 
Compared with state-of-the-art schemes, PATH achieves $4.7-7.8\times$ higher throughput, $>14.5\times$ lower tail latency, and $>61.4\%$ fewer memory accesses under insertions, laying a scalable foundation for next-generation data-centric computing.
}


\maketitle

\section{Introduction}\label{sec1}
\begin{figure*}
\centering
  \includegraphics[]{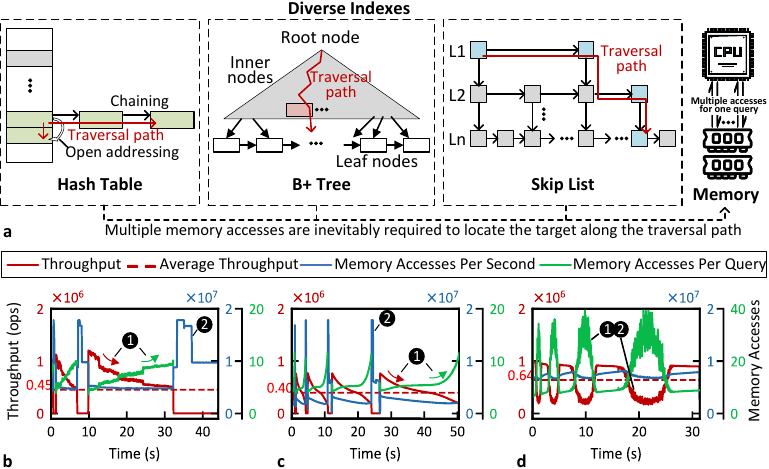}
\vspace{-15pt}
\caption{\textbf{Multiple memory accesses and processes are required along the traversal path for diverse indexes.} \textbf{a}, An illustration of the multiple memory accesses required by diverse indexes along the traversal path to locate the target. \textbf{b-d,} Visualizing the performance issues with representative hash indexes. \textbf{b} \textit{PCLHT} \cite{lee2019recipe} and \textbf{c}, \textit{Level Hashing} \cite{zuo2018write}: \myCircled[]{1}increased real-time memory accesses per query and reduced real-time throughput due to \textit{collisions}, and \myCircled[]{2}massive memory accesses during \textit{resizing}. \textbf{d}, The performance of a representative extendible hash index that adopts incremental resizing, namely, \textit{Dash} \cite{lu13dash}: \myCircled[]{1}\myCircled[]{2}fluctuating real-time memory accesses and throughput due to \textit{collisions and resizing}.}
  \label{fig.intro}
\end{figure*}

Data lies at the core element of information processing and modern artificial intelligence (AI) \cite{faulkner2020data, huawei2022general}.
Big data emerges with richer information content for processing and greater intelligence for AI.
While storing such massive data demands enormous storage capacity, what matters even more is how to efficiently organize and retrieve it.
Without effective data organization, even the largest data repositories become inaccessible and remain underutilized.
Indexes, as the essential data structures that bridge raw data and actionable information, serve precisely this purpose and constitute the fundamental backbone of modern storage systems.
However, the ever-widening gap between computation speed and memory access latency has made data movement the primary cost and bottleneck, known as the \textit{memory wall}. 
Specifically, the two major categories of indexes, namely hash-based indexes and tree-based indexes, are both constrained by the memory wall.
For hash indexes, inevitable hash collisions cause different keys to map to the same location, which forces queries to perform multiple rounds of memory accesses and comparisons to traverse hash buckets, regardless of the collision resolution strategy used (Fig.\ref{fig.intro}a). 
Meanwhile, as data volume continuously increases, hash tables must be expanded through resizing that incurs full-table rehashing and substantial data movement \cite{zuo2018write}.
Tree-based/like indexes, such as B+ Tree (or skip list), face a similar bottleneck: queries typically require node-by-node traversal from the root (Level 1) to a target leaf (Level n), introducing several dependent memory accesses along the traversal path. 
In addition, continuous insertions may trigger node splitting or tree rebalancing, further causing data movement and structural maintenance overhead.
Despite many efforts that have been made to improve the indexing schemes \cite{zuo2018write, lee2019recipe, lu13dash}, observations (Fig.\ref{fig.intro}b-d) reveal that state-of-the-art (SOTA) indexes, still suffer from significant performance issues (Supplementary Note 1 provides more thorough study).

Fortunately, beyond data storage, non-volatile memory (NVM), also referred to memristive memory, exhibits efficient in-situ computational capabilities within the memory arrays, providing a Processing-in-memory (PIM) solution to the issue.
Prior studies on PIM have already demonstrated broad potential in neural networks \cite{xiao2020analog, chi2016prime, buchel2025efficient, lin2025deep, leroux2025analog}, scientific computing \cite{le2018mixed}, and graph analytics \cite{song2018graphr}.
Specifically, the content-addressable memory (CAM) capability of memristive memory arrays, so-called ReCAM, enables in-situ data matching and thus provides an attractive basis for in-situ indexing where data resides.
However, it remains unclear how indexing can effectively benefit from PIM's unique capabilities.
The key challenge lies in the mismatch between the structure/operations of indexes and the hardware features of PIM.
More precisely, there is a gap between in-situ operations supported by PIM, such as matrix-vector multiplication (MVM) or CAM capabilities of memory arrays, and the operations required for indexing, i.e., Insert, Search, Update, and Delete (ISUD). 
Moreover, maintaining the structure of indexes needs fussy control logic, and the limitations of integrated circuit manufacturing processes render its implementation on PIM uneconomical.
In this work, we propose PATH, a PIM architecture which paves a new path for the first time to build in-situ indexing. 
PATH first reformulates the fundamental indexing operations into highly parallel in-situ CAM operations executed directly within memory arrays.
Due to the inherent parallelism of the in-situ ISUD, PATH creates new opportunities for indexing algorithms to adopt simplified hash table designs and flatter tree structures.
Taking hash indexes as a representative example, conventional hash indexes adopt tailored bucket and various collision resolution strategies to maximize the load factor before table resizing, at the cost of increased memory accesses and higher process latency.
PATH breaks the inherent trade-off among memory accesses, load factor, and process latency in conventional hashing schemes through hardware/software (HW/SW) co-design. 
By adopting ultra-large logical buckets and in-memory moving, PATH virtually eliminates the cost of hash collision resolution and significantly reduces resizing overhead.
Moreover, the co-design can be extended to accelerate a wide range of indexes on top of PATH, laying a scalable foundation for next-generation data-centric computing.
\section{Results}\label{sec2}

\subsection{PATH architecture and in-situ ISUD abstraction}\label{overview}

\begin{figure*}
  \centering
  \includegraphics[width=\linewidth]{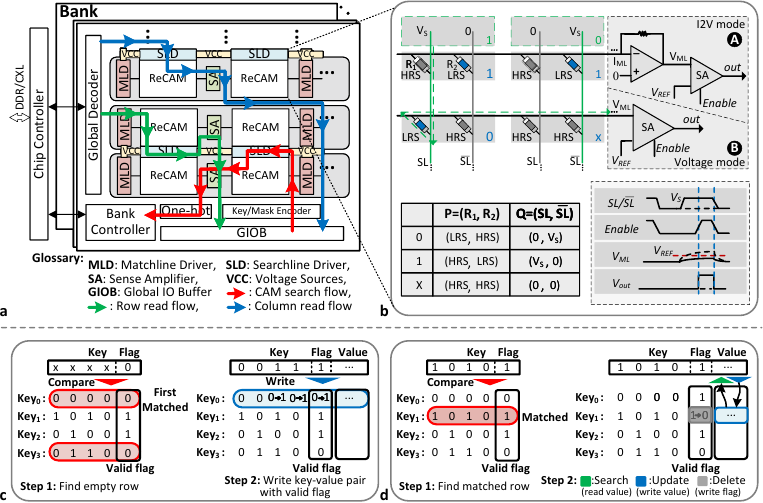}
    \vspace{-15pt}
  \caption{\textbf{Illustration of the PATH architecture and the in-situ ISUD.} \textbf{a,} PATH architecture. The green, blue, and red bold lines represent the directions of the data flow for the normal memory row read, column read, and for the CAM operation, respectively. \textbf{b,} ReCAM array and two kinds of sensing modes: \myCircled[]{A} current-to-voltage (I2V) mode and \myCircled[]{B} voltage mode. \textbf{c,} In-situ insert operation. \textbf{d,} In-situ search, update, and delete operations.}
  \label{fig.arch}
\end{figure*}

As shown in Fig.\ref{fig.arch}a,  PATH adopts a conventional memory-like architecture and employs only minimal logic to support ReCAM operations, except for storage.
Thus, PATH avoids complex interconnections (e.g., network-on-chip), costly auxiliary circuits of the previous PIM architectures used for neural networks \cite{chi2016prime, song2017pipelayer}.
PATH consists of several banks, each containing numerous ReCAM arrays, as illustrated in Fig.\ref{fig.arch}b, to support both data storage and ReCAM operations.
Specifically, one ReCAM bit is composed of a pair of memristive cells, each having two resistance states: high resistance state (HRS) and low resistance state (LRS). 
The stored data $P$ is encoded by the resistance combination of these two cells, where `0', `1', and `X' correspond to LRS-HRS, HRS-LRS, and HRS-HRS states, respectively.
During search, the query data $Q$ is applied as voltage pulses ($0/V_S$) on the search lines. 
When $Q$ and $P$ mismatch, $V_S$ is applied to the LRS cell, so the ML voltage is pulled up to a high value that is larger than the reference voltage ($V_{REF}$). 
Otherwise (matched), $V_S$ is applied to the HRS cell, maintaining the ML at a low voltage.  
The chip controller is responsible for the overall control of conventional memory access commands and our newly added PIM commands, and forwards them to the bank.
Each bank is independently controlled by the bank controller. 
As depicted, Fig.\ref{fig.arch}a shows the data flows for three modes: normal read (row read), CAM, and column read (symmetric to row read).
Supplementary Note 2 provides a more detailed design.
Fig.\ref{fig.arch}b illustrates the two sensing modes for CAM operations: current-to-voltage (I2V) mode and voltage mode.
The impact of cell/array parameters, such as array size, sensing mode, and ON/OFF ratios, on the sensing margin is discussed in Methods.

Diverse indexes adopt hash table, tree, or other structures to guide queries to a local area that contains the target data, such as a hash bucket in hash tables or a leaf node in B+ trees.
They then perform insert, search, update, or delete operations in that area. 
If the target area cannot accommodate so many data items, a structural maintenance process is triggered, such as hash collision resolution, table resizing, or tree node splitting, to move data and maintain the structure. 
Therefore, we first bridge the fundamental functional gap between indexes, namely ISUD in a local area, and the computational capability of PIM by introducing an in-situ ISUD abstraction for ReCAM arrays, which enables the ISUD to be performed in situ within the arrays. 
Fig.\ref{fig.arch}c--d illustrate the proposed in-situ ISUD operations. 
We store keys and corresponding valid flags at each row of the ReCAM array, which enables parallel comparison between the input (key, flag) and stored ones.
Since values do not need to take part in the comparison, each bit of them is stored in the same row as the corresponding key (or another array with the same row ID) using one normal cell rather than one ReCAM cell (i.e., two cells).
The \textit{insert} operation, as shown in Fig.\ref{fig.arch}c, has two sub-steps: first, we use the ReCAM array's parallel comparison capability to find a row with the valid flag of 0 (i.e., an empty row), and then we write the KV items and valid flag into that row.
If multiple empty rows are found, we simply select one row to do the second step, and here the first empty row is selected.
For the \textit{search}, \textit{update} and \textit{delete} operations, the combination of the input key and a valid flag of 1 is fed into ReCAM for comparison.
Thus, a matching row indicates a valid row that holds the same key.
As shown in Fig.\ref{fig.arch}d, according to the operation type, we then need to read and return the value for \textit{search}, write the new value for  \textit{update}, or clear the valid flag for \textit{delete}.

\begin{figure*}
  \centering
  \includegraphics[width=0.93\linewidth]{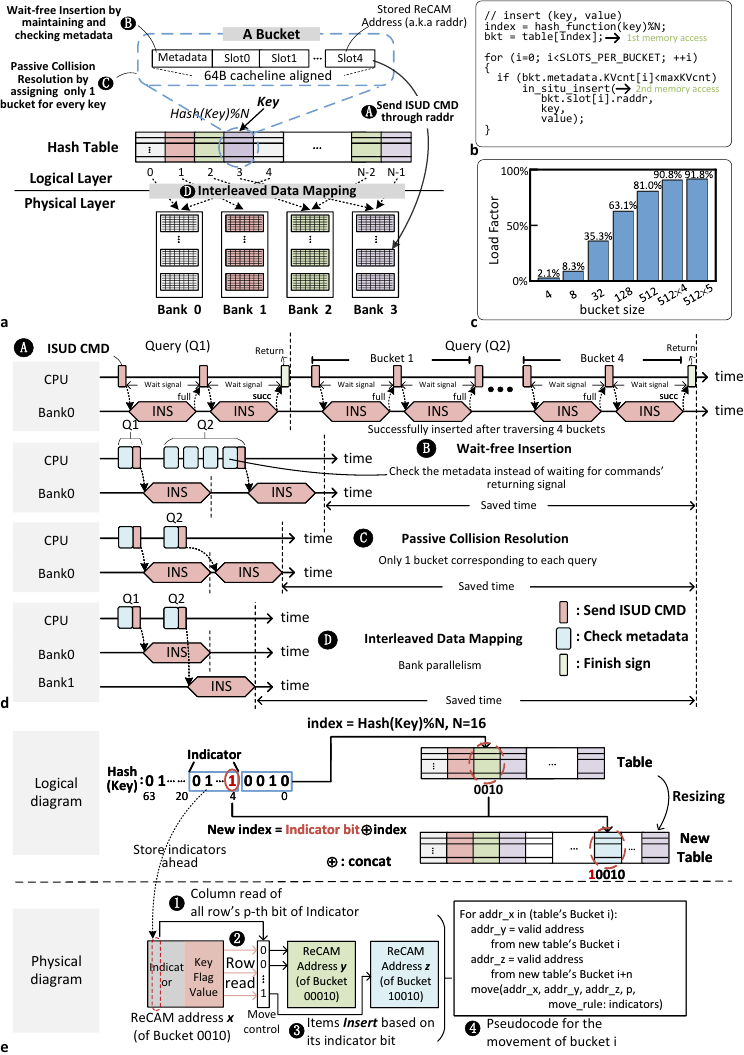}
  \caption{\textbf{The HW/SW co-designs.} \textbf{a}, An illustrative co-designed hash index. \textbf{b}, Pseudocode for the access optimization strategy, i.e., wait-free insertion and passive collision resolution. \textbf{c}, The load factors for different bucket sizes ($B_{size}$) under tests with 100 million KV items under 0.99 skewness. \textbf{d}, The processing time for proposed co-design strategies. \textbf{e}, In-memory moving for resizing. The processing time illustration assumes each bucket has only 2 slots for simplicity, and two insert queries to be processed. For the basic case, we assume a linked-list of 4 buckets for active collision resolution.}
  \label{fig.hwsw}
\end{figure*}

\subsection{Hardware/Software Co-design}\label{hwsw}
With the inherent parallelism of the in-situ ISUD, PATH creates new opportunities to restructure the indexing algorithm. 
Here, we propose HW/SW co-designs for hash indexes as a representative example to break the inherent trade-off among memory accesses, load factor, and process latency in conventional hashing schemes. 
\begin{equation}\label{eq1}
\begin{aligned}
  T_{conv} &\approx T_{hash} + \overbrace{(N_{mem}-1)}^{\text{mismatched accesses}}\times\overbrace{(\underbrace{T_{mem}}_{\myCircledWhite[0.5]{1}} + T_{cmp})}^{\text{per-access overhead}}+\overbrace{(T_{mem}+T_{op})}^{matched\ access},\ N_{\text{mem}} \approx \left\lceil \frac{N_{probe}}{n_{conv}} \right\rceil \\
  T_{path} &\approx T_{hash} + (N_{mem}^{'}-1)\times(\underbrace{T_{mem}}_{\myCircledWhite[0.5]{2}}+T_{isud}^{miss})+(T_{mem}+T_{isud}^{match}),\ N_{\text{mem}}^{'} \approx \left\lceil \frac{N_{probe}}{n_{path}} \right\rceil 
\end{aligned}
\end{equation}
We use Eq.~\ref{eq1} to characterize the latency of conventional and PATH-based hash indexes, which serves as a guideline for designing HW/SW co-designs to improve PATH.
For conventional case, the total process latency $T$ is approximated as the sum of hash computation time $T_{hash}$, the cost of mismatched accesses (memory access time $T_{mem}$ + CPU comparison time $T_{cmp}$), and the final successful matched access overhead ($T_{mem}$ + operation time $T_{op}$). 
Here, $N_{mem}$ denotes the number of memory accesses needed, which is approximated as $\left\lceil N_{\text{probe}} / n_{conv} \right\rceil$, where $N_{\text{probe}}$ is the number of KV items that need to be probed and $n_{conv}$ is the number of KV items that can be read in a single memory access.
For the PATH case, the cost model is similar, except that the mismatched and matched access costs are represented by \(T_{isud}^{miss}\) and \(T_{isud}^{match}\), respectively. 
The two cases differ in two fundamental ways. 
First, in conventional schemes, the memory access \myCircledWhite[0.6]{1} targets a hash bucket that contains KV items, after which the CPU performs the comparisons. 
In PATH, however, the memory access \myCircledWhite[0.8]{2} targets a bucket that contains only \emph{addresses} of the ReCAM array, i.e., a \textbf{logical bucket}; the subsequent process is carried out by issuing an in-situ ISUD command to the corresponding address (Fig.\ref{fig.hwsw}a($\myCircled[]{A}$)). 
Second, due to the high parallelism of ISUD in PATH, specifically, 512 KV items are processed in parallel in situ using a 512-row array, whereas only 4--8 KV items can be fetched into CPU caches for serial or limited parallel processing in conventional schemes \cite{zuo2018write, lee2019recipe}. 
Thus, $n_{path} \gg n_{conv}$, which implies $N_{mem}' \ll N_{mem}$, and the total processing latency is expected to satisfy $T_{path} < T_{conv}$. 
To further improve efficiency, we adopt ultra-large logical buckets and make the PATH-based hashing scheme cache-friendly by storing multiple ReCAM addresses in each bucket, e.g., five slots per bucket with one address per slot, as shown in Fig.\ref{fig.hwsw}a.
With this organization, each logical bucket fits within a 64B cacheline, so a single memory access can prefetch the whole bucket into the CPU cache. 
Although the logical bucket itself is small, it logically covers a large amount of data by pointing to multiple ReCAM arrays.
Increasing the bucket size in conventional hash indexes does not help to reduce the process latency, since $N_{mem}$ is only determined by the number of probes and the number of KV items that can be fetched in one memory access.
Importantly, our ultra-large logical bucket does not incur additional space overhead, because the hash table size remains the same as in conventional schemes.

We then propose optimization co-designs to (1) reduce $T^{*}_{isud}$ latencies in PATH by wait-free operations, especially for insertions, (2) restrict $N'_{mem}$ to 1 by passive collision resolution; with both (1) and (2), we thereby improve the final latency $T_{path}$, and (3) parallelize queries across multiple banks by interleaved data mapping.

\textbf{Wait-free insertions.}
Specifically, this strategy is built on a key mechanism: unlike the \textit{search} operation, which must wait for data to be returned, the \textit{insert} operation can be executed without waiting as long as sufficient space is available to accommodate the inserted KV item.
To achieve wait-free \textit{insertion}, we maintain \textit{metadata} within each bucket (Fig.\ref{fig.hwsw}a($\myCircled[]{B}$)) that tracks the count of valid KV items for each associated ReCAM array.
Crucially, our compact metadata design ensures that the entire bucket structure fits entirely within a single 64-byte cacheline.
If the count is less than the array's row size, it signifies a definite success of insertion without any waiting required.
In this case, we perform a wait-free \textit{insert} to improve efficiency, yielding $T_{insert}^{miss}$ and $T_{insert}^{match}$ that are both approximately 0.
Thus, the entire process consists of only two sub-steps (Fig.\ref{fig.hwsw}b): 1) check the metadata in the bucket to find a ReCAM array with space, and 2) perform a wait-free insert.

\textbf{Passive collision resolution.}
Conventional hash indexes \cite{zuo2018write,lee2019recipe} usually only hold 4--8 KV items per bucket.
During continuous insertions, a bucket will soon be filled up, necessitating collision resolution.
If \textit{resizing} is performed directly at this time, it results in extremely insufficient utilization of the hash table space, as the load factor may be very low , i.e., $2.1\%$, as shown in Fig.\ref{fig.hwsw}c.
Thus, various \textbf{active} strategies \cite{lee2019recipe, zuo2018write, DBLP:conf/usenix/ChenHDZ20} are introduced to resolve collisions before the final \textit{resizing} and improve the load factor, such as chaining, open addressing, double hashing, or space arrangement by moving KV items.
Through a series of active collision resolution strategies, they achieve a maximum load factor of around $50\%$-$90\%$ before resizing at the cost of a big $N_{mem}$ that is more rounds of memory accesses and comparisons to traverse buckets.
However, the number of KV items that a bucket can hold ($B_{size}$) in PATH-based hashing schemes far exceeds that of conventional hash schemes, reaching $5\times 512$ (we have $5$ ReCAM arrays for one bucket, and each array can hold $512$ KV items).
This allows us not to take any active collision resolution strategies.
The \textit{resizing} operation is triggered directly once a bucket is full.
At that point, the table is also nearly full. 
Therefore, it is called \textbf{passive} collision resolution, as no active collision resolution action is taken before \textit{resizing}.
This is because the more KV items a bucket can hold, the more margin it naturally provides to tolerate short-term insertion imbalances between different buckets, thereby improving the final load factor.
As shown in Fig.\ref{fig.hwsw}c, the load factor increases as $B_{size}$ increases. 
We are able to obtain a load factor of $91.8\%$.
Due to passive collision resolution, we only have one location (i.e., one bucket) corresponding to a hashed key.
As a result, $N'_{mem}$ is thereby strictly limited to 1.
Thus, each insertion requires at most two memory accesses.
One is to retrieve the data of the target bucket, and the other is to send an ISUD command (see Fig.\ref{fig.hwsw}b).
For other types of queries, such as search, update, and delete operations, the number of memory accesses is also reduced since $N'_{mem}$ is limited to 1; however, the ISUD command may need to be attempted several times as one bucket contains multiple ReCAM addresses.

\textbf{Interleaved data mapping.}
To explore more parallelism, we allocate hash buckets in a bank-level interleaving manner.
As shown in Fig.\ref{fig.hwsw}a, assuming a 4-bank memory, the $i$-th bucket is mapped to the bank $i\%4$.
Specifically, the map here refers to all addresses stored in the $i$-th bucket coming from the bank $i\%4$.
Arrays in different banks can execute ISUD operations simultaneously, thereby improving the overall performance of wait-free or concurrent queries.
Increasing the number of banks can further improve performance.

By integrating the above strategies, a basic single-level PATH-based hashing scheme is illustrated in Fig.\ref{fig.hwsw}a, and its corresponding processing time is shown in Fig.\ref{fig.hwsw}d.
In the example, Query 1 (Q1) traverses two slots to find insertion space, while Q2 needs to traverse four buckets (2 slots per bucket) to find space. 
In contrast, with wait-free insertion ($\myCircled[]{B}$), we only need to check the metadata instead of waiting for the insert command's returning signal. However, we still need 4 times metadata check (corresponding to 4 buckets) for Q2. With passive collision resolution ($\myCircled[]{C}$), each insertion only needs 1 metadata check. 
Furthermore, the interleaved mapping strategy ($\myCircled[]{D}$), allows commands to execute simultaneously across different banks.

\textbf{Rule-guided in-memory data moving.} For indexes, another critical aspect is structure maintenance, such as hash table resize or tree node split, both requiring substantial data movement. 
Therefore, we design a rule-guided memory movement mechanism to perform the movement directly in memory and avoid expensive off-chip memory accesses. 
This design is motivated by our observations of hash table resize and split operations, whose movements are highly regularized. 
Specifically, during hash table resize, the data in each bucket either remains in place or moves to a new position, and that position is determined by one bit of the hash value, i.e., single-bit check. 
During tree node split, each data item either remains in place or moves to a new position, and that position is determined by the comparison result between the key of the data item and the pivot key, i.e., simple comparison. 
Therefore, we can use these rules as guidance to design the memory movement mechanism, enabling data movement for structure maintenance to be completed directly in memory.

We still take hash indexes as an example to illustrate the rule-guided in-memory moving mechanism.
When resizing, the hash index will be recalculated via the following equation:
\begin{equation}
  \begin{aligned}
    index = hash(key)\%N \rightarrow index=hash(key)\%(2N) \nonumber
  \end{aligned}
\end{equation}
where $N$ is the hash table size, and we assume $N$ is the power of 2.
As illustrated in Fig.\ref{fig.hwsw}d, when $N=16$, the lowest 4 bits represent the index.
The next effective bit of $hash(key)$, i.e., the $4$-th bit, serves as the resizing \textit{indicator}, indicating whether the target KV item in $index$-th bucket should move to $index$-th bucket or $(index+N)$-th bucket in the new hash table.
In the example, the current indicator bit is $1$, so the new index equals $1\oplus index$ (here, $x\oplus y$ refers to concat(x, y), i.e., $1\oplus 0010=10010$, binary) or $index+N$ (i.e., $2+16=18$, decimal).
We select the next 16 bits of $hash(key)$ as the complete indicator.
This suffices to indicate the following 16 resizes, equivalent to expanding the space by $65536\times$.
By storing index indicators in the ReCAM array with corresponding KV items, we can conduct the \textit{resizing} directly in memory via a HW/SW co-design.
In hardware, a move control is introduced in the bank controller to move data in a ReCAM array based on the index indicator.
The process consists of three steps, as shown in the physical diagram of Fig.\ref{fig.hwsw}e.
$\myCircled[]{1}$ A column read obtains the currently used indicator bits of all rows into the registers (namely, REG\_Indicator),
$\myCircled[]{2}$ A traversal row read retrieves every KV item stored in each row of the ReCAM array,
and $\myCircled[]{3}$ Insert the KV items into their new location based on the respective indicator bit in REG\_Indicator.
The column read is not necessarily required (optional), because traversal row reads can read out the indicator bits simultaneously. 
The movement will not be cross-bank to keep the control logic and data flow simple.
Thus, the new $(index+N)$-th bucket should be assigned the same bank as the original $index$-th bucket.
In software, as pseudocode in $\myCircled[]{4}$ shows, we can use the in-memory moving function for an individual ReCAM to complete movement of an entire bucket and further the entire table.
Here, $move(addr_x, addr_y, addr_z, p)$ encapsulates the hardware in-memory moving function, representing moving the KV items in the ReCAM array at address $addr_x$ to address $addr_y$ if its indicator bit ($p$-th bit) is $0$, otherwise to the ReCAM array at address $addr_z$.
The in-memory moving function, can combine with different hashing schemes, e.g., the split operation of EH-based hashing schemes, since our scheduling unit is a ReCAM array, which does not conflict with hash structural designs. 
In summary, the rule-guided in-memory moving is shown in hash indexes as indicators. 
When applied to tree-based indexes, the rule becomes a simple comparison, as tree node splitting is usually determined by comparing the key with the pivot key. 
Thus, across diverse indexes, the rules are simple and general, making them easy to implement in hardware and flexibly apply to various indexes.

\begin{figure}[t!]
  \centering
  \includegraphics[]{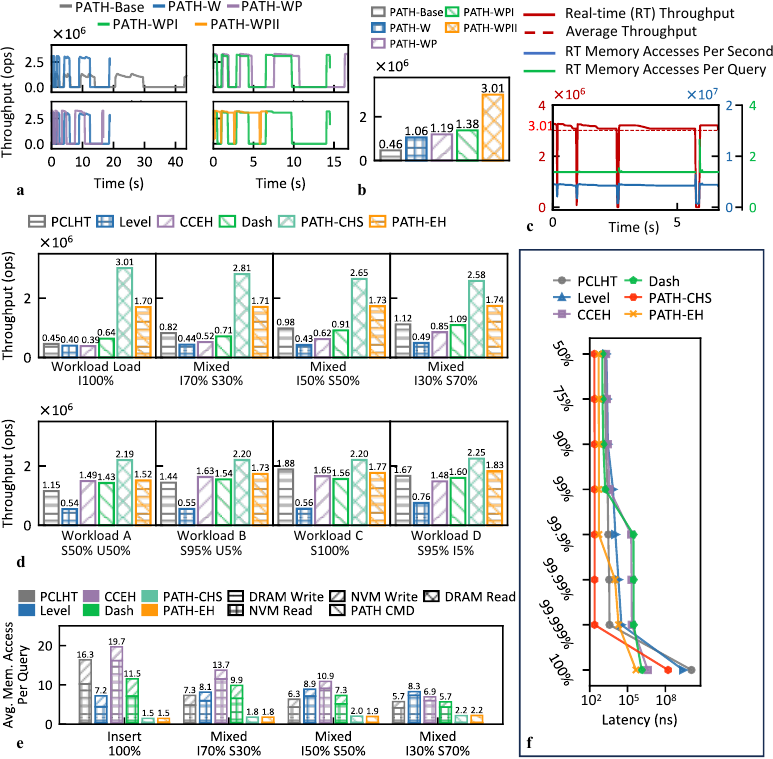}
  \caption{\textbf{Comprehensive experimental results.} \textbf{a-b}, The contribution of each component in our HW/SW co-design to (a) real-time throughput and (b) average throughput under insertions. \textbf{c}, Real-time throughput and memory accesses of our PATH-CHS design. \textbf{d}, Average throughput under different workloads. \textbf{e}, Comparison of average memory accesses per query. \textbf{f}, Tail latency comparison for insertions. 
  We implemented the PATH architecture in the system-level simulator Gem5 \cite{binkert2011gem5}.
On this basis, we instantiated two variants of hashing schemes (i.e., PATH-CHS and PATH-EH) alongside the state-of-the-art hash indexes (i.e., PCLHT \cite{lee2019recipe}, Level \cite{zuo2018write}, CCEH \cite{nam2019write}, and DASH \cite{lu13dash}).
PATH-CHS is a PATH-based conventional hashing with a single-level table, as illustrated in Fig.\ref{fig.hwsw}a.
Similar to PCLHT and Level, PATH-CHS has blocked resizing. 
PATH-EH is the incorporation of our HW/SW co-designs into CCEH.
  \textit{PATH-Base} serves as the baseline design, employing in-situ ISUD with multiple ReCAM addresses per bucket and a linked list of length 4 to resolve collisions;
\textit{PATH-W} builds on PATH-Base by incorporating \textbf{\underline{W}}ait-free insertion;
\textit{PATH-WP} extends PATH-W with \textbf{\underline{P}}assive collision resolution;
\textit{PATH-WPI} further enhances PATH-WP by adding \textbf{\underline{I}}nterleaved mapping;
\textit{PATH-WPII} finally incorporates \textbf{\underline{I}}n-memory moving into PATH-WPI.}
  \label{fig.evaluations1}
\end{figure}

\subsection{System-Level Performance Evaluation}\label{subsec_eval}
First, we investigate the effect of individual techniques introduced in \S\ref{hwsw} under workload Load, i.e., continuous insertions.
The real-time throughput in Fig.\ref{fig.evaluations1}a shows the each part's contribution.
Wait-free insertions speed up both the insertion and resizing, as resizing also contains continuous insertions. 
Passive collision resolution and interleaved mapping further improve performance, while in-memory moving reduces the resizing latency sharply.
As a result, we improve the performance efficiency by $6.5\times$ compared to PATH-base, as shown in Fig.\ref{fig.evaluations1}b.
Fig.\ref{fig.evaluations1}c illustrates the real-time throughput and memory accesses of PATH-CHS under insertions, which can be compared with Fig.\ref{fig.intro}b-d, showing that PATH-CHS achieves significantly better and more stable performance.
More real-time test results under different workloads are shown in Supplementary Note 4.

Subsequently, we comprehensively evaluated the holistic performance of the PATH-based hashing schemes under realistic workloads in Fig.\ref{fig.evaluations1}d.
Specifically, PATH-CHS achieves $6.6\times$, $7.6\times$, $7.8\times$, and $4.7\times$ faster for average insertion throughput (workload load) compared to PCLHT, Level, CCEH, and Dash. 
And PATH-based CCEH (PATH-EH) outperforms the original CCEH $4.4\times$ for average insertion throughput.
As the insertion ratio decreases, the performance of other schemes improves, while our scheme's performance declines.
This is because the wait-free operation is only valid for insertion.
In those low insertion ratio cases, PATH-CHS is still $2.3\times$ and $1.3\times$ faster than PCLHT under $30\%$ and $5\%$ insertion ratio, respectively. 
On average, we only require 1.5 memory accesses per query, as shown in Fig.\ref{fig.evaluations1}e.
Supplementary Note 5 provides more performance results under multi-thread execution.
The tail latency of different percentiles is shown in Fig.\ref{fig.evaluations1}f.
For schemes that have a blocked resizing, PATH-CHS achieves $14.5\times$, $15.5\times$, $72.2\times$ lower latency than PCLHT, and $78.6\times$, $135.2\times$, $15.9\times$ lower latency than Level, at 99.99th, 99.999th and 100th percentiles of insertions.
For EH-based hashing schemes, PATH-EH achieves $20.7\times$, $11.3\times$, $8.6\times$ lower latency than CCEH, and $29.9\times$, $15.8\times$, $2.8\times$ lower latency than Dash, at 99.99th, 99.999th, and 100th percentiles of insertions.
By adopting splitting rather than full-table resizing, PATH-EH achieves better latency at 100th percentiles than PATH-CHS at the cost of worse latency at other percentiles since it has more complex processing flow.
Overall, our design can be integrated with diverse hashing schemes to further improve their performance. 
We present a sensitivity analysis of PATH under different NVM write latencies in Supplementary Note 6, demonstrating the applicability of PATH.
The area and energy evaluation of PATH are shown in Supplementary Note 7, and the comparison of PATH with near-data processing (NDP) approaches is shown in Supplementary Note 8.
We further extend our design to the B+ tree, namely PATH-tree, and the performance comparison is shown in Supplementary Note 9.

\section{Discussion}\label{sec_dis}
This paper presents PATH, a well-tailored processing-in-memory architecture for in-situ indexing.
Since we have ISUD directly realized in PIM, another perspective is to build a new index algorithm suitable for PIM rather than using the conventional indexes.
However, the algorithm established in this way will ultimately settle on known indexes such as hashing-based indexes or tree-based indexes.
This is because the scale of executing ISUD is limited to a single array, which is not enough for indexing the whole data.
You can utilize multiple ReCAM arrays to implement larger-scale ISUD, but activating more arrays simultaneously incurs significant energy costs.
For example, to realize a direct ISUD of $100$ million KV items, it is necessary to activate more than $100$ thousand arrays simultaneously, assuming an array size of $1024$, which is unacceptable.
Therefore, you should guide queries to either one ReCAM array or a specific number of parallel ReCAM arrays.
This is consistent with conventional schemes that guide queries to a single tree node (with tree-based indexes) or a hash bucket (with hashing-based indexes).
This paper uses various hash indexes as the primary example and also discusses tree-based indexes, such as B+ Tree, to demonstrate the generality of the proposed approach. 
It is expected to extend to a wide range of index types in future work.

\section{Methods}\label{sec11}

\subsection{Experimental Setup}\label{sec.setup}

\textbf{Environment.} 
All experiments are performed on the platform with two Intel(R) Xeon(R) Silver 4314 CPU (@2.4GHZ) and eight 32 GB DDR4 2666MT/s DRAM DIMMs.
The operating system is Ubuntu 18.04.1 with Linux kernel version 5.4.0.
The GCC version 7.5.0 is used for compilation. 
We conduct experiments using the Gem5 simulator \cite{binkert2011gem5}.
The Gem5 simulator is a modular simulation platform for computer-system architecture.
Gem5 supports NVM controller and NVM memory natively.
On top of this, we build a mock PMDK library that allows hashing schemes that are performed on NVM using the intel PMDK library \cite{intelpmdk}, to run on Gem5 with minimal modifications.
We add our PATH architecture.
We adopt CPU and cache parameters similar to those in \cite{prasad2021memristive}. 
The simulated processor is a 32-core X86 architecture running at 2 GHz with a write pending queue size of 128. 
The cache hierarchy includes a private 32 KB L1 instruction cache that is direct-mapped and has a 2-cycle latency, a private 32 KB L1 data cache that is 4-way set associative with LRU replacement and has a 2-cycle latency, and a shared 8 MB L2 cache that is 16-way set associative with LRU replacement and has a 20-cycle latency. 
We use DDR4-2400 DRAM and NVM as the main memory. 
A standard default DRAM configuration is used in Gem5. 
The DRAM main memory configuration consists of 8 GB DDR4-2400 DRAM with 1 channel, 1 rank per channel, and 8 banks per rank, and its timing parameters are tRCD-tCL-tWR-tWTR: 14-14-15-5 ns. 
For NVM, it is derived from 8 GB DDR4-2400 DRAM, and the read ($tREAD$) and write ($tWR$) latencies are set to typical values of 20 ns and 100 ns, respectively \cite{ielmini2016resistive}. 
Our PATH is also derived from DDR4-2400 memory.
Though PATH can serve as main memory, for the convenience of comparison, PATH and NVM-type main memory are separated.
The same 20ns delay as $tREAD$ is used for our ReCAM operation, i.e., $tCAM$.
Here, we do not use aggressive ReCAM operation parameters from the ReCAM prototypes \cite{li20131, lin20167, chang2016designs, garzon20234}.
Instead, we use parameters similar to the conventional NVM to make a fair comparison, which means that performance improvement does not come from faster operations of our PATH.
All experiments are conducted using the same simulation parameters in Gem5.

\textbf{Evaluated Hashing Schemes.}
The following representative hashing schemes are evaluated:

$\bullet$ \texttt{PCLHT}: PCLHT \cite{lee2019recipe} is an NVM version of conventional CLHT hashing schemes \cite{DBLP:conf/asplos/DavidGT15} with linked list based cache-friendly buckets.

$\bullet$ \texttt{Level}: Level hashing \cite{zuo2018write} is the origin of Level-based hashing schemes with a multi-level structure for cost-efficient resizing. 

$\bullet$ \texttt{CCEH}: CCEH \cite{nam2019write} is developed based on extendible hashing (EH) \cite{ellis1983extendible}, i.e., the NVM version of EH-based hashing, which adopts split to extend table size when needed.

$\bullet$ \texttt{Dash}: Dash \cite{lu13dash} is an enhanced version of CCEH \cite{nam2019write}. It introduces several techniques, such as bucket load balance and optimistic locking, for better load factor and performance. 
These baselines were compared against our proposed solutions: \texttt{PATH-CHS}, representing our blocked resizing implementation on the PATH architecture, and \texttt{PATH-EH}, representing our directory-based dynamic resizing implementation.

\textbf{Benchmark.}
For all workloads, we warm up the hash table with 1 million items, and then we execute 20 million operations.
We adopt mixed workloads with different insertion ratio to illustrate our PATH.   
Besides, we also use the real-world workloads from Yahoo Cloud Serving Benchmark (YCSB) \cite{cooper2010benchmarking} workloads A-D.
All workloads are generated in the Zipfian distribution with 0.99 skewness as in \cite{wang2023seph,lu13dash}.

\subsection{Software-Hardware Support}\label{sec.sup}
To back our PATH architecture, this section discusses some software and hardware support.

\textbf{Software Support.}\label{interface}
To enable efficient utilization of our in-situ operations, PATH provides user-space application programming interfaces (APIs) similar to previous PIM systems \cite{prasad2021memristive, chi2016prime}.
We allow developers to: (1) allocate and free memory arrays in PATH, (2) normal read and write to the memory array, and (3) execute in-situ ISUD and in-memory data moving.
We present exemplary code using our APIs in Supplementary Note 3.
Since PATH improves only the fundamental ISUD operations to the bucket, the concurrent control logic of concurrent operations to different buckets is the same as the original software hashing schemes.
Our API library offers simple but sufficient controls over the proposed PATH, and it also enables users to adopt these interfaces to accelerate other data indexes in the future.

\textbf{DIMM and Multi-DIMM Support.}
To further scale capacity, it is feasible to organize multiple PATH chips under a dual in-line memory module (DIMM) \cite{standard2016ddr4} when the data cannot fit within a single chip.
Further, a system can include multiple DIMMs to store more data.
The parallelism provided by multiple chips/DIMMs can be used to enhance the performance of concurrent queries similar to multi-bank parallelism (interleaved mapping mentioned in \S\ref{hwsw}).
The in-memory moving function will only occur in the same bank (see \S\ref{hwsw}).
Since we have guided each query to a specific ReCAM array, there is no need for communication between multiple chips/DIMMs.
There are no restrictions on the operations to each array, and an array can respond to read/write or ReCAM operations at any time.

  \begin{figure*}
  \centering
  \includegraphics[]{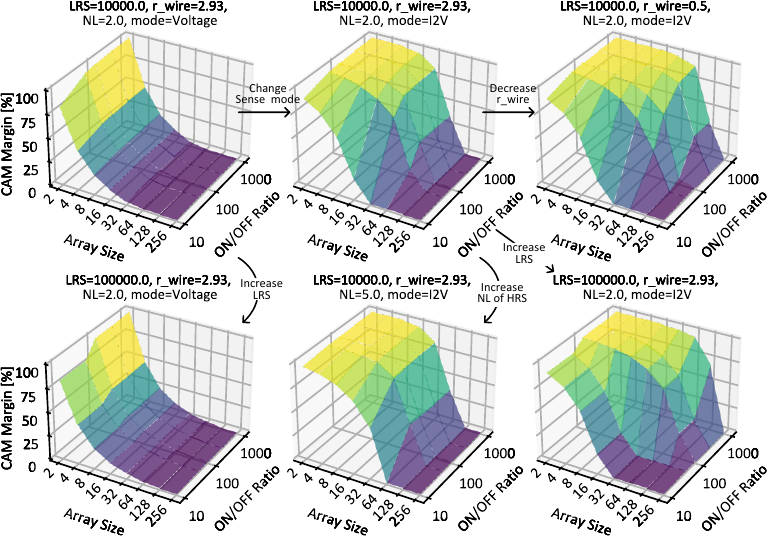}
  \caption{\textbf{The CAM sensing margin under different cell/array parameters,} including array sizes, sensing modes, ON/OFF ratios, and nonlinearity (NL) of HRS. Here, the array size N refers to $N\times N$ ReCAM array.} 
  \label{fig.margin}
\end{figure*}

\textbf{Cell and Array support.}
Here, we further explore the cell/array parameters to trade off between CAM sense margin and performance. 
The default cell is modeled with a linear LRS resistance ($R_L$) of $10k\Omega$ and a nonlinear HRS resistance ($R_H$) of $r\times R_L$ with a nonlinearity NL ($@1.5V$). 
Here, r is the $R_H/R_L$ (ON/OFF) ratio and the nonlinearity of the cell is defined as $NL=2\times(R(V/2)/R(V))$ \cite{wu2019reram}. 
The sensing voltage $V_s$ is set to $0.4V$ and write voltage of cell is set to $1.5V$.
The default $NL$ is set to $2$ which means a linear HRS.
The default interconnect resistance ($r_{wire}$) between adjacent cells in the array is set to $2.93\Omega$ \cite{wu2019reram, yang2022memory}.
Conventional ReCAM array directly senses the voltage on the ML \cite{yang2022memory}, as shown in Fig.\ref{fig.arch}b($\myCircled[]{B}$), and the margin degrades rapidly (Fig.\ref{fig.margin}) because the effective resistance drops quickly under parallel connection. 
By virtually grounding the ML and then converting current to voltage for sensing (Fig.\ref{fig.arch}b($\myCircled[]{A}$)), this degradation can be suppressed as the array scales (Fig.\ref{fig.margin}).
In sum, we draw the following conclusions from the SPICE simulation results: 1) the I2V sensing mode can improve the margin and thus support larger arrays; 2) increasing the NL of HRS and the ON/OFF ratio can both improve the margin, but only when the margin is initially nonzero, and the achievable improvement is bounded; and 3) increasing LRS or decreasing $r_{wire}$ can improve the margin in I2V mode, whereas voltage mode cannot.
Thus, we adopt the I2V sensing mode.
To further improve parallelism, we adopt multiple arrays, such as n arrays, to extend CAM functionality vertically, that is, allowing more data to be compared. 
We do not extend it horizontally, because the matching bit length of the key is typically limited (64-bit key in the index is usually sufficient).
The key and flag are sent to the n arrays in parallel, and the matching results from the arrays are then sent to a two-step one-hot module to obtain the matching row number.
The first step is a standard one-hot module that outputs a match row for the CAM array, and the second step gather the results from n arrays to output the final match row number.
In this way, we decouple the one-hot module for multiple arrays and support a configurable two-step one-hot module, enabling different values of n.
In this paper, we adopt CAM arrays of size 128 because the sensing margin is sufficient, and we set n to 4 by default to form 512-row CAM arrays.
More discussions on the consistency, reliability, and endurance are provided in Supplementary Note 10.

\backmatter


\bibliography{sn-bibliography}

\end{document}


\large Supplementary Information
\\
\vspace{-10pt}
\noindent\rule{\linewidth}{0.4pt}
\vspace{-10pt}
\begin{center}
\LARGE
\textbf{In-situ Indexing via Memristive Content-Addressable Memory}
\end{center}

\noindent\rule{\linewidth}{0.4pt}

\setcounter{tocdepth}{1}

\tableofcontents
\clearpage

\section{Memory Issues for Indexes}\label{background}
Non-volatile memory deployed as main memory provides high-capacity, non-volatile, low-latency, and high-throughput data storage \cite{bittman2019optimizing, wang2023seph}.
The capacity and performance advantages of NVMs make them particularly attractive for in-memory applications \cite{yang2020empirical}.
Researchers have developed carefully-tailored indexes to better utilize the potentials of NVM, including tree- and hashing-based indexes, such as wB+-Tree \cite{chen2015persistent}, BzTree \cite{arulraj2018bztree}, and Level Hashing \cite{zuo2018write}.
Tree-based indexes typically have a time complexity of $O(logN)$ for point queries, where $N$ represents the amount of stored data, and support effective range queries.
By contrast, hashing-based indexes can provide fast point queries with a time complexity of $O(1)$, so they are widely adopted by in-memory systems \cite{garcia1992main, fan2013memc3, wang2023seph} where point queries are dominant.
As a result, many hashing schemes are proposed for NVM, such as Level Hashing \cite{zuo2018write}, CLevel Hashing \cite{DBLP:conf/usenix/ChenHDZ20}, CCEH \cite{nam2019write}, and Dash \cite{lu13dash}.

\subsection{The Memory Access Issues for Hash Indexes}\label{nvm_hash}

Different hashing schemes have their specific structural designs to optimize the performance.
Based on how the \textit{resizing} operation is performed, the recent SOTA hashing schemes for NVM can be generally categorized into three groups: 1) \textbf{Conventional hashing}, exemplified by PCLHT \cite{lee2019recipe}, which requires a \textit{full table rehashing} for resizing, 2) \textbf{Level-based hashing}, such as Level hash \cite{zuo2018write} and CLevel hash \cite{DBLP:conf/usenix/ChenHDZ20}, which employs a \textit{multi-level} structure for cost-efficient (1/3 of the table) resizing, and 3) \textbf{EH-based hashing}, including CCEH \cite{nam2019write} and Dash \cite{lu13dash}, which adopts \textit{incremental resizing} (limited to a segment, usually comprising hundreds of KV items) from extendible hash (EH) \cite{ellis1983extendible}.

Conventional and Level-based hashing schemes usually have higher throughput due to relatively simpler structural hierarchy \cite{zuo2018write}, while EH-based hashing schemes have better tail-latency and scalability due to incremental resizing \cite{wang2023seph}.
Different hashing schemes have their own superiority and are therefore suitable for different scenarios \cite{huang2023indexing}.
While existing studies have made remarkable progress in enhancing the hashing schemes, hash indexes still suffer from suboptimal performance due to 1) more memory accesses and operations (\S\ref{mot1}) caused by hash collisions, and 2) costly resizing overhead (\S\ref{mot2}).
\begin{figure*}[b]
  \centering
  \includegraphics[width=1.1\linewidth]{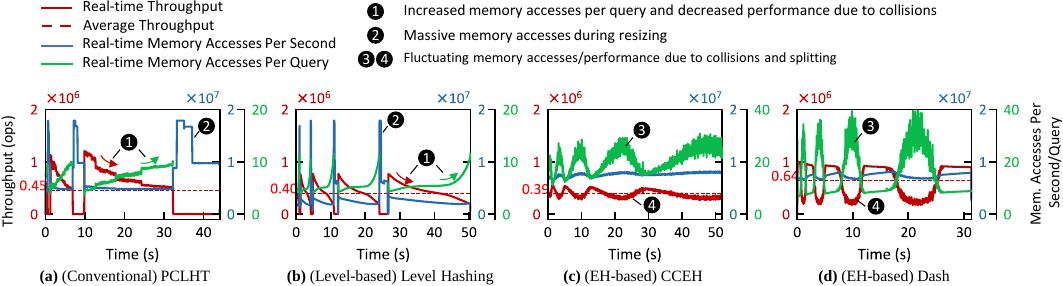}

  \caption{Real-time throughput (red line), memory accesses per second (blue), and memory accesses per query (green) under insertions for (a) Conventional, (b) Level-based, and (c-d) EH-based hashing schemes. Since (a) and (b) adopt a blocked resizing, memory accesses per query are not recorded during resizing.} 
  \label{fig.mot}
\end{figure*}

\subsubsection{Increased Number of Memory Accesses and Operations}\label{mot1}
The existing hashing schemes adopt a cache-friendly table design, so that a bucket can be accessed at most with only one memory access \cite{zuo2018write, wang2023seph, lee2019recipe}.
Hash collisions drive them to use strategies like linear probing, chaining, or multiple locations per hashed key, for improved space efficiency (i.e., load factor) before resizing.
This results in multiple rounds of memory accesses and operations for a single query.
The CPU cache is not very helpful in reducing the memory accesses due to its small capacity compared with the large amount of KV items \cite{zhang2015mega}.
The increased number of memory accesses and operations leads to a comprehensive decline in performance.
Fig.\ref{fig.mot} validates the performance cost brought by collisions under different hashing schemes.
It can be seen in Fig.\ref{fig.mot}(a-b) that the real-time throughput (shown in red line) declines as the number of memory accesses per query (shown in green line) increases caused by collisions, and in Fig.\ref{fig.mot}(c-d) that the real-time throughput fluctuates as the number of memory accesses per query fluctuates.
In summary, hash collisions are one of the primary reasons for degraded performance, since they incur more memory accesses/operations.


\subsubsection{Resizing Overhead}\label{mot2}
When the hash table cannot resolve the current hash collision, a \textit{resizing} operation will be performed to expand the table capacity.
Level-based hashing schemes \cite{zuo2018write, DBLP:conf/usenix/ChenHDZ20} reduce the amount of data movement for resizing to one-third of the table.
EH-based hashing schemes \cite{nam2019write, lu13dash,wang2023seph} adopt incremental resizing, i.e., split, to move only a segment (typically hundreds or thousands of KV items) or even one-third of a segment.
However, such data movement is still expensive, leading to poor performance efficiency and long tail-latency.
As shown in Fig.\ref{fig.mot}(a) and (b), resizing incurs a sudden drop in real-time throughput (indicated by the red line) as well as massive memory accesses (blue line).
The EH-based hashing scheme, as shown in Fig.\ref{fig.mot}(c-d), adopts \textit{split} to avoid one-time massive data movement of resizing, however, the cost is shared by normal queries, which brings drastic fluctuations in memory accesses and throughput. 


\subsection{The Memory Access Issues for Tree-based Indexes}
\begin{figure*}[h]
  \centering
  \includegraphics[]{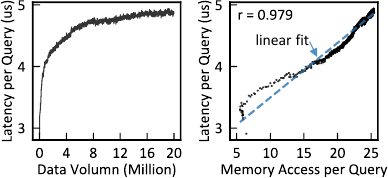}

  \caption{The memory issues in B+ tree (uTree).} 
  \label{fig.mot-b-tree}
\end{figure*}

We take uTree (a kind of B+ Trees) as an example of tree-based indexes to illustrate the memory access issues.
The B+ Tree organizes data in a hierarchical tree structure, where leaf-node stores the actual data and non-leaf nodes store the keys for indexing.
These tree-based indexes have a time complexity of $O(logN)$ for point queries since they need to traverse from the root to the leaf node.
Fig.\ref{fig.mot-b-tree} (left) shows that, as the data volume increases, query latency increases accordingly. 
Fig.\ref{fig.mot-b-tree} (right) further shows the strong linear correlation (pearson correlation coefficient $r=0.979$) between query latency and memory access.

\clearpage
\section{PATH Architecture}\label{detailed architecture}
\subsection{Architecture}
\begin{figure}[h]
  \centering
  \includegraphics[]{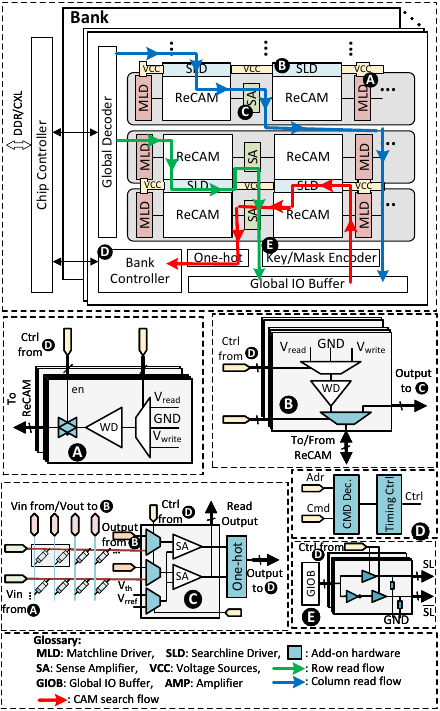}
  \caption{Architecture and peripheral circuit.}
  \label{fig.detailed_arch}
\end{figure}

As shown in Fig.\ref{fig.detailed_arch}, The chip controller (CTRL-C) is responsible for the overall control of conventional memory access commands and our newly added PIM commands coming from the interface.
Byte-addressable DDR or CXL interfaces will both work.
Specifically for a bank, it's controlled by the bank controller (CTRL-B, $\MyCircled[]{D}$).  
The green and red arrows illustrate the data flow for a normal read operation (e.g., row read) and ReCAM operation.
Since the column read operation is needed in accelerating resizing, its data flow is also shown. 
The transmission gates are added to the matchline driver (MLD, $\MyCircled[]{A}$) to close MLs when performing in-situ ReCAM operations, while allowing currents to pass when performing normal memory operations.
The searchline driver (SLD, $\MyCircled[]{B}$) is responsible for supplying voltages to the ReCAM array for both ReCAM and normal memory operations, as well as forwarding the results to SAs when performing column reads. 
Two different reference voltages for read ($V_{rref}$) and ReCAM ($V_{th}$) operations are required for SAs ($\MyCircled[]{C}$).
We share MLD/SLD/SAs between two adjacent arrays to reduce the area.
For ReCAM operations, each input data bit is encoded by the Key/Mask module ($\MyCircled[]{E}$) to generate two signals for SL and $\textoverline{SL}$, respectively.
And the result from SAs is sent to the global IO buffer (GIOB) for a normal read, or to the one-hot module for a ReCAM operation to get the matching row number and then is further forwarded to the bank controller to perform the next operation.

\clearpage

\section{Application Programming Interface (API) and Exemplary Code}\label{code-example}
\begin{figure}[h]
  \centering
  \includegraphics[]{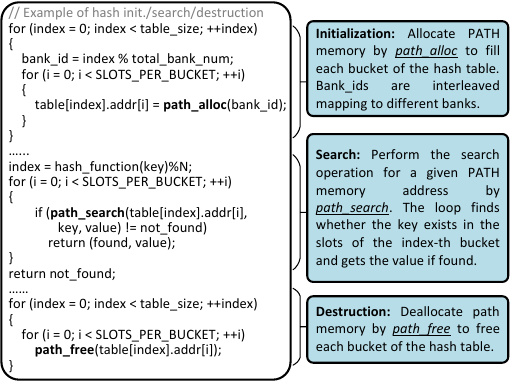}
  \caption{Example code snippet.}
  \label{fig.interface}
\end{figure}

To enable efficient utilization of our in-situ operations, PATH provides user-space application programming interfaces (APIs) similar to previous PIM systems \cite{prasad2021memristive, chi2016prime}.
We allow developers to: (1) allocate and free memory arrays in PATH, (2) normal read and write to the memory array, and (3) execute in-situ ISUD and in-memory data moving.
We show an example code snippet to illustrate how our designed APIs initialize the hardware and execute the data queries for hash indexes in Fig.\ref{fig.interface}, respectively.
We illustrate three basic functions, i.e., \texttt{path\_alloc()}, \texttt{path\_free()}, and \texttt{path\_search()}. 
The argument, i.e., $bank_{id}$, of \texttt{path\_alloc()} controls the allocation of a ReCAM array from a specific bank.
\texttt{path\_alloc()} returns an address to users to freely use this memory region.
In the example, \texttt{path\_search()} will use these allocated PATH addresses to accomplish the corresponding search operations.
At last, we call \texttt{path\_free()} to de-allocate the memory region

\clearpage

\section{Real-time Throughput and Memory Accesses}\label{appedix_eval_sec}
\begin{figure}[h!]
  \centering
  \includegraphics[]{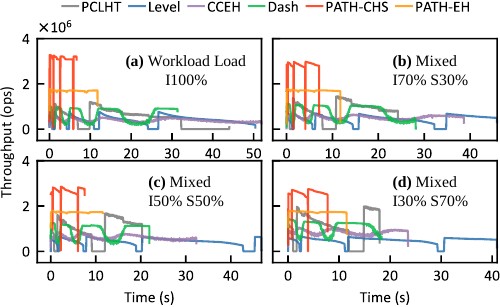}
  \caption{Real-time throughput of all schmems.}
  \label{fig.real_all}
\end{figure}

\begin{figure}
  \centering
  \includegraphics[]{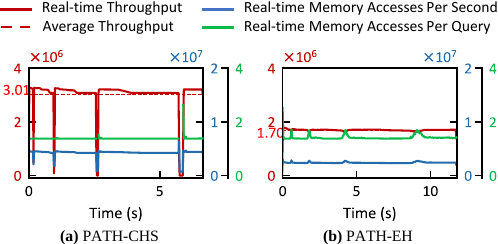}
  \caption{Real-time throughput and memory accesses of PATH-CHS and PATH-EH under insertions.}
  \label{fig.ops_my}
\end{figure}

Fig.\ref{fig.real_all} shows the real-time throughput of all schemes under different insert/search ratio, and Fig.\ref{fig.ops_my} illustrates the real-time memory accesses of our PATH-based hashing schemes under insertions, which can be compared with Fig.\ref{fig.mot} to show that our schemes require far fewer memory accesses. 
It can be preliminarily observed that the PATH-based hashing schemes achieve the highest throughput and exhibit the least fluctuation.

\clearpage

\section{Multi-thread Test}

\begin{figure}[h!]
  \centering
  \includegraphics[]{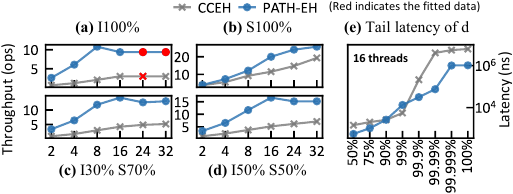}
  \caption{Multi-thread test.}
  \label{fig.multi}
\end{figure}

Given that EH-based schemes (e.g., CCEH) typically exhibit superior scalability compared to PCLHT, we focus this evaluation on comparing PATH-EH against CCEH. 
Fig. \ref{fig.multi} demonstrates that PATH-EH achieves significant performance gains over this state-of-the-art baseline under multi-thread execution.
In addition to the testing of 100\% insertion, we reduce the amount of test data by $10\times$ for other multi-thread tests to speed up the evaluation.
PATH-EH outperforms CCEH $5.1\times$, $3.1\times$, $3.1\times$, and $3.1\times$ for the average insertion throughput under 8, 16, 24, and 32 threads, respectively. 
Note that some tests are failed due to the instability of Gem5 when running multiple threads, so the data was fitted (shown in red).

The possible reason why the PATH-based hashing scheme does not scale much better than the original hashing scheme under multiple threads is that the scalability is affected by the concurrency control and structure of the hashing scheme itself. 
Since we have accelerated the query process by PATH, the concurrency control and structure of the hashing scheme may have a greater impact in this situation.
In short, the faster the hardware is, the higher the overhead ratio of the software control.
We believe that our PATH will bring new opportunities for hash table design with extremely lightweight control.
Nevertheless, PATH-EH is still $3.1\times$ faster than CCEH when both reach their performance bounds.
And we also achieve a better tail latency, as shown in Fig.\ref{fig.multi}e.
PATH-EH achieves $5.4\times-53.6\times$ lower latency at 99.99th, 99.999th, and 100th percentiles than CCEH.

\clearpage

\section{Sensitivity to NVM Write Latency}

\begin{figure}[h!]
  \centering
  \includegraphics[]{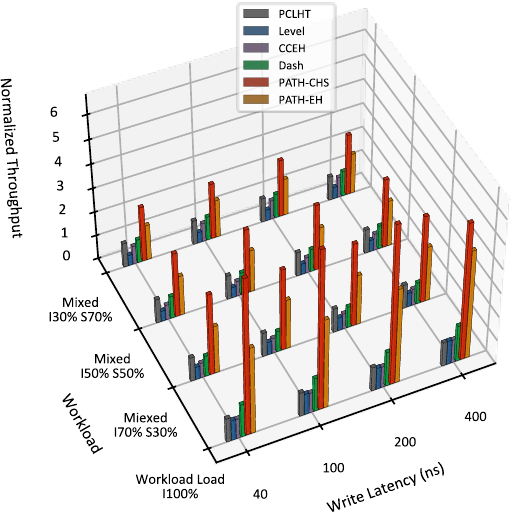}
  \caption{Sensitivity to NVM write latency.}
  \label{fig.nvm_sensitivity}
\end{figure}

We evaluate the sensitivity of PATH-based hashing schemes to NVM write latency by varying the write latency from 40 ns to 400 ns. 
Across all write latencies, PATH-based hashing schemes consistently outperform the original hashing schemes, as shown in Fig.\ref{fig.nvm_sensitivity}. 
Under 100\% insertions, the relative improvement of PATH-CHS over PCLHT varies across latencies, but the difference remains small, with a maximum deviation of only 14.05\% from the relative improvement observed at the default 100 ns write latency. 
Under the other workloads, the variation is even smaller, with maximum deviations of 9.29\% for PATH-CHS over PCLHT and 10.96\% for PATH-EH over CCEH compared to the relative improvement at the default 100 ns write latency.

\clearpage

\section{Area and Energy}
As shown in Table \ref{tab:PATH}, the total area of a single PATH bank is 40.88 $mm^2$, dominated by the arrays, which occupies 35.73 $mm^2$. 
This indicates that the memory array remains the primary contributor to the overall area cost. 
As mentioned in \S\ref{detailed architecture}, we share the MLD/SLC/Sense between two adjacent arrays, so their numbers are half of the array number.
The peripheral circuits, including MLD, SLD, and Sense, account for 1.44 $mm^2$, 2.09 $mm^2$, and 1.61 $mm^2$, respectively, while CTRL-B and Key/Mask introduce only negligible overhead. 
Importantly, the total area of our added components is 13.23 $mm^2$, corresponding to only 4.05\% of the total PATH area.

\begin{table}[htbp]
    \centering
    \caption{PATH Configurations}
    \label{tab:PATH}
    \begin{tabular}{|c|c|c|c|}
        \hline
        \textbf{Component} & \textbf{Area ($mm^2$)} & \textbf{Params.} & \textbf{Spec.} \\
        \hline
        \hline
        \multicolumn{4}{|c|}{\textbf{Bank properties}}\\
        \hline
        CTRL-B\textsuperscript{1} & 4.95e-3 & Total & 1 \\ \hline
        \multirow{3}{*}{XB Array\textsuperscript{*}}  & \multirow{3}{*}{35.73}& Bits per Cell & 1 \\ 
        & & Size & 128$\times$130\textsuperscript{\(2\)}\\
        & & Total & 524288 \\
        \hline
        MLD\textsuperscript{*} & 1.44 & Total & 262144$\times$128 \\ \hline
        SLD\textsuperscript{*} & 2.09 & Total & 262144$\times$130 \\ \hline
        Sense\textsuperscript{3*}  & 1.61 & Total & 262144$\times$130 \\ \hline
        Key/Mask\textsuperscript{*}  &  1.60e-6 & Total & 65 \\ \hline
        Bank Total & 40.88& Size & 1GB\textsuperscript{4} \\ \hline
        \hline\hline
        \multicolumn{4}{|c|}{\textbf{PATH properties}}\\
        \hline
        Our Added Total & 13.23 & Area Percentage & 4.05\%\\ \hline
        PATH Total & 327.05 & Size & 8GB (8 banks) \\ \hline
    \end{tabular}
    \footnotesize\textsuperscript{1}The area of CTRL-B is obtained through Design Compiler at 65nm and scaled to 32nm. It includes the global buffer, the one-hot module, and the control of composite commands (in-memory moving and in-situ ISUD).\\
    \footnotesize\textsuperscript{2}The array has 130 columns, where 128 are for normal data and 2 additional columns are used for flag bit.\\
    \footnotesize\textsuperscript{3}The Sense module includes components such as SA and Mux. 
    \footnotesize\textsuperscript{4}The total size excludes the additional flag bits.\\
    \footnotesize\textsuperscript{*}The area of the array and peripheral circuits are modeled using NeuroSim \cite{peng2019dnn+, chen2018neurosim} based on the 32nm technology node, with a minimal cell size of $4F^2$.

\end{table}

\begin{figure}[h]
  \centering
  \includegraphics[]{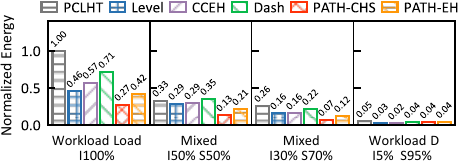}
  \caption{The energy consumption comparisons under different workloads.} 
  \label{fig.energy}
\end{figure}

Fig.\ref{fig.energy} shows the energy consumption of all schemes under different workloads. 
We adopt the NVM energy model from Aliens \cite{wu2018aliens, wu2019low}.
DRAM energy model and parameters are obtained from \cite{vogelsang2010understanding}.
The energy consumption of PATH-based hashing schemes is significantly lower than that of the original hashing schemes, with reductions of 73\% for PATH-CHS over PCLHT and 26\% for PATH-EH over CCEH under 100\% insertions. 
As the insertion ratio decreases, the energy savings gradually diminish, but they remain positive at 30\% insertions. 
When the insertion ratio drops to 5\%, our scheme shows a slight increase in energy consumption compared with the traditional schemes, although the difference is small. 
The energy savings come from the trade-off between the reduced number of operations enabled by PATH's high-parallelism comparison capability and the higher per-comparison energy cost. 
This is because writes incur high energy cost. Under write-intensive workloads, the high-parallelism comparison capability of PATH reduces the number of operations and thus provides an energy-efficiency advantage. However, as the writing ratio decreases, the higher energy cost of each comparison caused by greater parallelism becomes more apparent, and the energy-efficiency advantage becomes less pronounced.
 
\clearpage

\section{Comparisons with NDP-based Indexes}\label{vsupmem-sec}
\begin{figure}[h]
  \centering
  \includegraphics[width=0.5\linewidth]{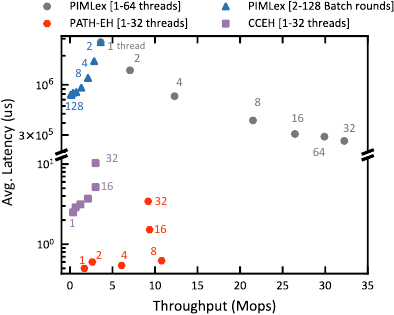}
  \caption{Comparisons with PIMLex under insertions.}
  \label{fig.upmem}
\end{figure}

UPMEM \cite{upmem, hyun2024pathfinding} is a general-purpose near-data-processing (NDP) memory architecture. 
Several works have focused on building indexes on UPMEM, e.g., PIM-Trie \cite{kang2023pim}, PIM-Tree \cite{kang2022pim-tree}, and PIMLex \cite{cui2025pimlex}.
These UPMEM-based indexes enhance throughput by processing batches of queries simultaneously in a bulk-synchronous fashion. 
However, this approach significantly degrades latency, rendering it highly unsuitable for latency-sensitive or quality of service(QoS)-sensitive applications, particularly for databases, operating systems, and real-time analytics, where latency is one of the most critical metrics.  
In this section, we compare PATH-EH with PIMLex \cite{cui2025pimlex} (the SOTA UPMEM-based index), to highlight the advantages of PATH.
The evaluation platform for UPMEM is a Linux server with two Intel(R) Xeon(R) Silver 4216 CPUs. For memory devices, the server is equipped with 4 conventional DRAM DIMMs (64GB per DIMM, 3200MT/s, DDR4) and 8 UPMEM DIMMs. 
Each UPMEM DIMM has 128 NDP modules. 
Thus, compared with PATH-EH, PIMLex is executed on a more powerful CPU and a greater number of processing units, comprising a total of 1024 NDP modules.
We adopt the same test parameters as those used in PIMLex \cite{cui2025pimlex}.
The test results for YCSB workload load (100\% insertion) are shown in Fig.\ref{fig.upmem}.

By default, PIMLex processes all test data in 2 rounds of batches. Opting for a smaller batch size (i.e., increasing the number of processing rounds) can enhance the average latency performance, as shown in blue triangles in the figure. However, this comes at the cost of reduced throughput.
The increase in the number of threads can enhance both latency and throughput of PIMLex (shown in gray circles), as it results in a smaller batch size and leverages the parallel processing capabilities of multiple threads.
However, once the number of threads hits 32, any further increase in the number of threads leads to a drop in throughput.
It can be observed that, regardless of whether the number of rounds or the number of threads is increased, once the average latency drops to 0.78 seconds and 0.26 seconds respectively, it cannot be further reduced.
UPMEM is not designed with a domain-specific focus on indexing, lacking dedicated data pathways and processing hardware tailored for indexing tasks. 
This results in the average latency remaining at a very high level, being $315,088\times$ slower than conventional software indexing CCEH (in the best-case scenario with a single thread).
Thus, UPMEM-based indexes focus on maximizing the throughput by the numerous NDP modules with batch processing model.
The application domains for UPMEM-based indexes differ entirely from those of conventional software and PATH-based indexes, which are limited to latency-insensitive applications.
As previously introduced, with in-situ ISUD and a series of HW/SW co-design techniques, PATH can be adapted to various hashing schemes and PATH-based indexes achieve comprehensive improvements in both latency and throughput.
The reason why the maximum throughput of PATH-EH falls short compared to PIMLex is primarily due to the overhead of software locks, as well as the design choice to prioritize a simple and scalable NVM architecture that can easily expand to large capacities (only minimal circuit and 8 parallel processing units, i.e., 8 banks, compared to 128 NDP modules per DIMM for UPMEM). 

\clearpage

\section{PATH-Tree}\label{path-tree}
Benefiting from our in-situ ISUD and the extensibility of the design, we also implement PATH-Tree, the PATH-based B+ tree via our HW/SW co-designs.
Here, we simply replace the leaf nodes with ReCAM arrays and use in-situ ISUD to perform the corresponding operations. 
When a leaf node splits, a pivot key is selected randomly from the leaf, and splitting is performed by in-memory data movement guided by a key comparison rule.
Compared with SOTA conventional B+ trees, including Fast\&Fair \cite{hwang2018endurable}, uTree \cite{chen2020utree}, and NBtree \cite{zhang2022nbtree}, our approach achieves $1.5-2.9\times$ throughput under different workloads.
PATH-Tree also achieves $3.2-8.4\times$ lower latency than uTree, and $1.6-7.0\times$ lower latency than NBTree, at 99th to 100th percentiles of insertions.

\begin{figure}[h]
  \centering
  \includegraphics[]{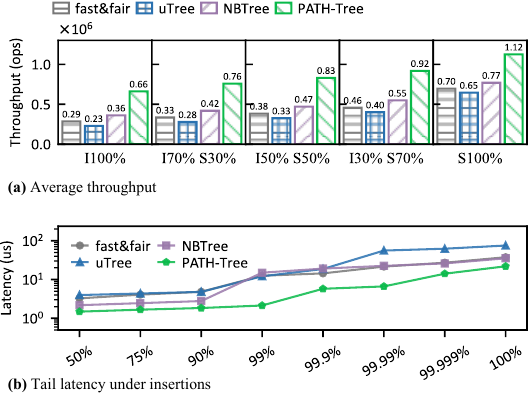}
  \caption{PATH-Tree compared with conventional B+ trees.}
  \label{fig.path-tree}
\end{figure}

\clearpage
\section{Discussion about Consistency, Reliability, and Endurance}\label{discuss}

\subsection{Consistency} 
No inconsistency will occur in PATH-based hashing schemes against crashes due to:
  1) 
  PATH guarantees in-situ ISUD commands to be atomic in the memory. 
  The modification of the metadata, i.e., the count of valid KV items for each slot, only needs to perform an atomic store.
  2) The in-memory moving transfers data from the old array to the new one, which is used to replace the item-by-item data movement during resizing/splitting in original hashing schemes. 
  If a crash occurs during moving, we can use the uniqueness of keys to delete duplicate data from the new array, restoring it back to a consistent state.
  The whole resizing or split operation can be protected using the original hashing schemes' methods, such as log-free, log, or Copy-on-Write (CoW).  
  For example, we can use the count of valid KV items for each slot as the same effect of the token used in Level hashing for log-free resizing \cite{zuo2018write}. 

\subsection{Reliability and Endurance.}

NVM technologies also have some faults \cite{xu2015impact, feinberg2018making}, such as retention failure, stuck-at-fault, and write disturbance.
Thus, PATH can adopt periodic refresh \cite{zhang2018cacf} and Error Correct Code (ECC) \cite{yoon2011free, xu2015impact, moon2020error} to ensure storage reliability similar to conventional NVM. 
As normal reads and writes are performed in a row manner, the periodic refresh and ECC should also be maintained in a row manner.
Particularly, an additional column-wise ECC should be maintained for index indicators since they are read in a column manner when resizing.
As for computing, i.e., ReCAM operations, there are dedicated ECC techniques \cite{krishnan2008error, roth2024error} for CAM memories and analog CAM (ReCAM) memories. 
In the experiments of this work, both conventional NVM and our PATH did not include these additional reliability overheads to demonstrate a basic performance comparison.
As for the limited endurance of NVMs, the natural even distribution of the hash function helps to evenly share writes out to different space, thus improving lifetime.
Assuming an 8-GB NVM with a typical endurance of $10^8$ write cycles per cell and an insertion rate of 10 Mops/s, a lifetime exceeding 20 years can be achieved. 
And for continuous update queries of the same key (extreme attack behavior) can be directly solved by conventional wear-leveling works on NVM, 
e.g., remapping strategies \cite{zhou2024drctl, prasad2021memristive, xu2019efficient}.

\bibliography{sn-bibliography}